\documentclass[cqg,preprint]{revtex4}%
\usepackage{amsfonts}
\usepackage{amsmath}
\usepackage{amssymb}
\usepackage{graphicx}%
\begin{document}
\title{Smarr formula for BTZ black holes in general three-dimensional gravity models}
\author{Chao Liang}
\affiliation{School of Mathematics and Physics, China University of Geosciences, Wuhan
430074, China}
\author{Li Gong}
\affiliation{School of Mathematics and Physics, China University of Geosciences, Wuhan
430074, China}
\author{Baocheng Zhang}
\email{zhangbc.zhang@yahoo.com}
\affiliation{School of Mathematics and Physics, China University of Geosciences, Wuhan
430074, China}
\keywords{Smarr formula; BTZ black holes; 3D gravity}
\pacs{04.70.-s, 04.70 Dy, 04.60 Kz}

\begin{abstract}
Recent studies have presented the interpretation of thermodynamic enthalpy for
the mass of BTZ black holes and the corresponding Smarr formula. All these are
made in the background of three-dimensional (3D) general relativity. In this
paper, we extend such interpretation into general 3D gravity models. It is
found that the direct extension is unfeasible and some extra conditions are
required to preserve both the Smarr formula and the first law of black hole
thermodynamics. Thus, BTZ black hole thermodynamics enforces some constraints
for general 3D gravity models, and these constraints are consistent with all
previous discussions.

\end{abstract}
\maketitle

\section{Introduction}

Smarr formula \cite{ls72}, together with the first law of black hole
thermodynamics \cite{bch73}, plays an important role in black hole physics.
For black hole solutions with a non-vanishing cosmological constant, in order
to maintain the scaling relation of Smarr formula, the cosmological constant
is considered as thermodynamic pressure
\cite{cm95,cck00,bpd111,bpd112,bpd113,cgp11,gmk12}, and the black hole mass,
which is usually regarded as analogous thermal internal energy in black hole
thermodynamics, is interpreted as a gravitational analog of thermodynamic
enthalpy $H$ \cite{krt09}, i.e. $M=H=U+PV$, where $U$ is the internal energy
and the pressure $P$ is derived from the cosmological constant. This has been
investigated for many different situations (see Ref. \cite{kmt16} and
references therein), and the consistent expressions for homogeneous Smarr
formula and the first law have been constructed, i.e. for a $D$-dimensional
singly-rotating black hole,%
\begin{equation}
dM=TdS+\Omega dJ+VdP, \label{1fl}%
\end{equation}%
\begin{equation}
\left(  D-3\right)  M=\left(  D-2\right)  TS+\left(  D-2\right)  \Omega J-2VP,
\label{hsf}%
\end{equation}
where $T$ is its temperature, $S$ is its entropy, $J$ is its angular momentum,
$\Omega$\ is its angular velocity, $V$ is its thermodynamic volume, and
$P=-\frac{\Lambda}{8\pi}=\frac{\left(  D-2\right)  \left(  D-1\right)  }{16\pi
l^{2}}$ is its pressure.

Recently, such interpretation as presented in Eqs. (\ref{1fl}) and (\ref{hsf})
has also been investigated for lower-dimensional black holes under the
background of extended thermodynamic phase space, or \textquotedblleft black
hole chemistry\textquotedblright\ \cite{fmm15}. For three-dimensional (3D)
Banados-Teitelboim-Zanelli (BTZ) black holes \cite{btz92} which is an solution
of 3D Einstein field equation plus a negative cosmological constant, it has
been confirmed \cite{fmm15,wwd06} that Eqs. (\ref{1fl}) and (\ref{hsf}) can be
satisfied simultaneously by assuming the negative cosmological constant as a
variable thermodynamic parameter. It was found that the extension to the
charged BTZ black holes is direct \cite{fmm15}, but to ensure the Reverse
Isoperimetric Inequality \cite{cgp11,akm14} a new thermodynamic work term
associated with the mass-renormalization scale has to be introduced. Moreover,
the mass of BTZ black holes as the enthalpy of the spacetime was also found to
be valid even under the condition of quantum correction \cite{bpd111}.

It is realized that the preservation of Eqs. (\ref{1fl}) and (\ref{hsf}) was
only investigated under 3D general relativity (GR), however. 3D GR propagates
no physical modes (it has no local degrees of freedom) and can be formulated
as a Chern-Simons (CS) gauge theory \cite{hw89}. The attempt to construct
generalizations \cite{djt82,mb91,bht09,hrz12,bht13,bht14} of 3D GR that
propagate gravitons has been a revival of interest in past years. In these new
3D gravity models, some of the thermodynamic parameters for BTZ black holes
which almost solves all the 3D gravity models have to be modified, and the
modified parameters still satisfied the first law of black hole thermodynamics
and the integral mass formula \cite{zbc16} but without forcing the
cosmological constant into a variable. Of course, the previous Smarr-like
formula, i.e. $M=\frac{1}{2}TS+\Omega J$, does not satisfy the scaling
relation as presented in Eq. (\ref{hsf}). In this paper, we expect to
investigate the relations among the thermodynamic parameters under the
background of extended 3D gravity models, and see if there are the same
relation as that in Eqs. (\ref{1fl}) and (\ref{hsf}) when considering the
cosmological constant as a new variable thermodynamic parameter. Moreover, a
motivation to make such investigation is that there are the phenomena of
continuous phase transition \cite{col95,pl99,mlz14} for BTZ black holes in
general 3D gravity models.

The organization of this paper is as follows. First, we revisit the
interpretation of thermodynamic enthalpy for the mass of BTZ black holes and
the homogeneous Smarr formula in the background of 3D normal Einstein gravity
in the second section. In the third section, we extend these discussions to
the case of 3D exotic Einstein gravity. Then, we investigate these relations
systematically in general 3D gravity models and see if they are still valid,
and investigate the phase transition for this case in the fourth section.
Finally, we summarize our conclusion in the fifth section.\ \ \ \ \ \ \ \ \ \ \ \ \ \ \ \ \ \ \ \ \ \ \ \ \ \ \ \ \ \ \ \ \ \ \ \ \ \ \ \ \ \ \ \ \ \ \ \ \ \ \ \ \ \ \ \ \ \ \ \ \ \ \ \ \ \ \ \ \ \ \ \ \ \ \ \ \ \ \ \ \ \ \ \ \ \ \ \ \ \ \ \ \ \ \ \ \ \ \ \ \ \

\section{Normal BTZ black holes}

The two-parameter BTZ solution represents a rotating black hole in 3D
spacetime with a negative cosmological constant, and its metric takes the form%
\begin{equation}
ds^{2}=-N^{2}dt^{2}+N^{-2}dr^{2}+r^{2}\left(  N^{\phi}dt+d\phi\right)  ^{2},
\label{btz}%
\end{equation}
where $\phi$ is an angle with the period $2\pi$ as the identification of the
black hole spacetime. The functions $N^{2}$ and $N^{\phi}$ are
\begin{equation}
N^{2}=-8Gm+\frac{r^{2}}{\ell^{2}}+\frac{16G^{2}j^{2}}{r^{2}},N^{\phi}%
=\frac{4Gj}{r^{2}},
\end{equation}
where $G$ the 3D Newton constant. Its Killing horizons are found by setting
$N^{2}=0$; this gives
\begin{equation}
r_{\pm}=\sqrt{2G\ell\left(  \ell m+j\right)  }\pm\sqrt{2G\ell\left(  \ell
m-j\right)  },
\end{equation}
We may assume without loss of generality that $j\geqslant0$ and assume that
$\ell m\geqslant j$, to ensure the existence of an event horizon at $r=r_{+}%
$.$\ $Thus, the parameters $m$ and $j$ can be expressed as%
\begin{equation}
m=\frac{r_{+}^{2}+r_{-}^{2}}{8G\ell^{2}},\text{ \ \ }j=\frac{r_{+}r_{-}%
}{4G\ell}. \label{mj}%
\end{equation}
And the temperature $T$ and the angular velocity $\Omega$ take the values%
\begin{equation}
T=\frac{r_{+}^{2}-r_{-}^{2}}{2\pi r_{+}\ell^{2}},\text{ \ \ }\Omega
=\frac{r_{-}}{\ell r_{+}}%
\end{equation}
which are independent on the concrete 3D gravity models and present the
geometric properties for the 3D BTZ spacetime background. $\ $

The BTZ metric can be solved in the normal Einstein gravity with the
Lagrangian
\begin{equation}
L_{G}=\frac{1}{8\pi G}\left(  e_{a}R^{a}+\frac{\Lambda_{0}}{6}\epsilon
^{abc}e_{a}e_{b}e_{c}\right)  ,
\end{equation}
where the Lagrangians are given with 3-forms in which $e^{a}$ ($a=0,1,2$) is
the dreibein 1-forms, $\omega^{a}$ is Lorentz connection 1-forms, $\Lambda
_{0}=-\frac{1}{\ell^{2}}$ is the negative cosmological constant and
$R^{a}=d\omega^{a}+\frac{1}{2}\epsilon^{abc}\omega_{b}\omega_{c}$ is the
curvature 2-form field strengths. With the BTZ metric (\ref{btz}), the
thermodynamic parameters for normal BTZ black holes are
\begin{equation}
M_{G}=m,\text{ \ }J_{G}=j,\text{ \ }S_{G}=\frac{\pi r_{+}}{2G}.
\end{equation}
It is easy to confirm that the usual thermodynamic relations, $dM_{G}%
=TdS_{G}+\Omega dJ_{G}$ and $M_{G}=\frac{1}{2}TS_{G}+\Omega J_{G}$, hold in
this situation, but the mass formula obviously violates the scaling relation
required in the Smarr formula of rotating black holes. In order to rectify
this situation, it is possible to promote the cosmological constant to a
variable thermodynamic parameter, so that the corresponding relations from
Eqs. (\ref{1fl}) and (\ref{hsf}) can be reexpressed as%
\begin{align}
dM_{G}  &  =TdS_{G}+\Omega dJ_{G}+V_{G}dP,\nonumber\\
0  &  =TS_{G}+\Omega J_{G}-2V_{G}P,
\end{align}
where the pressure is $P\equiv\frac{1}{8\pi G\ell^{2}}$ and the conjugate
thermodynamic volume is $V_{G}=\frac{\partial M_{G}}{\partial P}=\pi r_{+}%
^{2}$ obtained by the mass $M\left(  S_{G},J_{G},P\right)  =\frac{4PS_{G}^{2}%
}{\pi}-\frac{\pi^{2}J_{G}^{2}}{128S_{G}^{2}}$. It is noted that the
thermodynamic volume is just the 3D geometric volume, independent on other
thermodynamic parameters except the entropy.

Like AdS black holes in higher dimensions \cite{krt09}, the mass of BTZ black
holes should be understood as thermodynamic enthalpy when the negative
cosmological constant is considered as a thermodynamic variable, but it is
noted that the thermodynamic enthalpy is zero here. This can be understood by
the fact \cite{km14} that the energy required to form the BTZ black hole is
balanced by the energy required by the external forces to place the black hole
into the cosmological environment.

\section{Exotic BTZ black holes}

The BTZ metric can also be solved in the exotic Einstein gravity \cite{ew88}
with the Lagrangian
\begin{equation}
L_{E}=\frac{\ell}{8\pi G}\left[  \omega_{a}\left(  d\omega^{a}+\frac{2}%
{3}\epsilon^{abc}\omega_{b}\omega_{c}\right)  -\frac{1}{\ell^{2}}e_{a}%
T^{a}\right]  ,
\end{equation}
where the torsion tensor is $T^{a}=de^{a}+\epsilon^{abc}\omega_{b}e_{c}$, but
the corresponding thermodynamic parameters becomes%
\begin{equation}
M_{E}=j/\ell,\text{ \ }J_{E}=\ell m,\text{ \ }S_{E}=\frac{\pi r_{-}}{2G}.
\label{ebtz}%
\end{equation}
This presents some exotic properties: the metric is the same as that of 3D
Einstein gravity but with reversed roles for mass and angular momentum, and
the entropy is proportional to the length of the inner horizon instead of the
event horizon. The corresponding understanding for these properties refer to
Refs. \cite{tz13,bz13}. It is easy to check that the usual thermodynamic
relations, $dM_{E}=TdS_{E}+\Omega dJ_{E}$ and $M_{E}=\frac{1}{2}TS_{E}+\Omega
J_{E}$, also hold in this situation.

As done for normal BTZ black holes, we also promote the cosmological constant
into a variable in this case to see if the Smarr formula can also obtained
like before. Based on this, now we express the mass of exotic BTZ black holes
as%
\begin{equation}
M_{E}=\frac{1}{\ell}\left(  \frac{2GS_{E}^{2}J_{E}}{\pi^{2}\ell}-\frac
{G^{2}S_{E}^{4}}{\pi^{4}\ell^{2}}\right)  ^{\frac{1}{2}}.
\end{equation}

When considering the pressure $P\equiv\frac{1}{8\pi G\ell^{2}}$, the conjugate
thermodynamic volume is obtained as%
\begin{align}
V_{E}  &  =\frac{\partial M_{E}}{\partial P}=\frac{\partial M_{E}/\partial
\ell}{\partial P/\partial\ell}\nonumber\\
&  =-\frac{3}{2\ell^{2}}\left(  \frac{2GS_{E}^{2}J_{E}}{\pi^{2}\ell}%
-\frac{G^{2}S_{E}^{4}}{\pi^{4}\ell^{2}}\right)  ^{\frac{1}{2}}+\frac{1}%
{2\ell^{2}}\left(  \frac{2GS_{E}^{2}J_{E}}{\pi^{2}\ell}-\frac{G^{2}S_{E}^{4}%
}{\pi^{4}\ell^{2}}\right)  ^{-\frac{1}{2}}\nonumber\\
&  =\frac{\pi}{2}\frac{3r_{+}^{2}r_{-}-r_{-}^{3}}{r_{+}}%
\end{align}
where the last step uses the relation in Eqs. (\ref{mj}) and (\ref{ebtz}). It
is noted that the exotic volume is not the geometric volume $V=\pi r_{+}^{2}$
and it also depends on the thermodynamic parameters $J_{E}$ and $P$ besides
the entropy $S_{E}$.

With the expressions of thermodynamic pressure and volume, we obtain the
relations again%
\begin{align}
dM_{E}  &  =TdS_{E}+\Omega dJ_{E}+V_{E}dP,\nonumber\\
0  &  =TS_{E}+\Omega J_{E}-2V_{E}P,
\end{align}
which presents the homogenous Smarr relation.

On the other hand, one might want to make the thermodynamic volume unchanged,
i.e. $V_{E}=V=\pi r_{+}^{2}$, whose thoughts stemmed from the geometric
expressions for the temperature and the angular velocity. Thus the Smarr
formula is written as
\begin{equation}
0=TS_{E}+\Omega J_{E}-2VP_{E},
\end{equation}
where a calculation leads to the pressure $P_{E}=\frac{3r_{+}^{2}r_{-}%
-r_{-}^{3}}{16\pi G\ell^{2}r_{+}^{2}}$. But unfortunately, this cannot
preserve the first law, i.e. $dM_{E}\neq TdS_{E}+\Omega dJ_{E}+VdP_{E}$, which
can be obtained only through the calculation of the derivative to the AdS
radius $\ell$. On the other hand, such expression for the pressure implies
that $P_{E}$ is not an independent variable relative to the entropy and the
angular momentum. This conflicts with the requirement of differential form of
the first law of black hole thermodynamics.

\section{BTZ black holes in general 3D gravity models}

As seen above for the normal and exotic BTZ black holes that the extensive
thermodynamic variables are model-dependent, so in general 3D gravity models,
the mass and the angular momentum are expressed as%
\begin{align}
M  &  =am+bj/\ell,\nonumber\\
J  &  =aj+b\ell m,\nonumber\\
S  &  =\frac{\pi}{2G}\left(  ar_{+}+br_{-}\right)  ,
\end{align}
where $a$ and $b$ are the parameters dependent on the concrete models, i.e.
for normal Einstein gravity, $a=1,b=0$; for exotic Einstein gravity,
$a=0,b=1$. If the parameters $a$ and $b$ are independent on the cosmological
constant or the AdS radius $\ell$, we can get the same expressions \
\begin{align}
dM  &  =TdS+\Omega dJ+VdP,\nonumber\\
0  &  =TS+\Omega J+2VP, \label{gsf}%
\end{align}
with the thermodynamic volume
\begin{equation}
V=\frac{2ar_{+}^{3}+3br_{+}^{2}r_{-}-br_{-}^{3}}{2r_{+}}\pi\ \label{gv}%
\end{equation}
which is not geometric, but will recover geometric volume when $a=1,b=0$. It
is also obvious that when $a=0,b=1$, $V=V_{E}$.

Then, in general 3D gravity models, are the parameters $a$ and $b$ independent
on the AdS radius? As well-known, BTZ black holes are the solutions almost for
all the 3D gravity models which are usually modified by topological terms,
i.e. topological massive gravity (TMG) \cite{djt82}, minimal massive gravity
(MMG) \cite{bht14}, which make the theory has one gravitational propagating
mode, or by higher-order derivative terms, i.e. new massive gravity (NMG)
\cite{bht09}, zwei-dreibein gravity (ZDG) \cite{bht13}, which make the theory
have two gravitational propagating modes, or by the two terms simultaneously,
i.e. general massive gravity (GMG) \cite{hrz12}, general minimal massive
gravity (GMMG) \cite{ms14}, or by other terms, i.e. Mielke--Baekler (MB) model
\cite{mb91}, et al. But for almost all these modified 3D gravity models,
either $a$ or $b$ or both of them are dependent on the AdS radius. Thus, we
have to check if the homogenous Smarr formula is still valid when taking
$a=a\left(  \ell\right)  ,b=b\left(  \ell\right)  $.

Assuming the Smarr formula in Eq. (\ref{gsf}) holds, we can obtain the same
expression for thermodynamic volume as in Eq. (\ref{gv}) but the parameters
$a$ and $b$ are now the function of AdS radius $\ell$. For the first law of
black hole thermodynamics, we have%
\begin{equation}
\frac{dM}{d\ell}=\frac{r_{+}^{2}+r_{-}^{2}}{8G\ell^{2}}\frac{da}{d\ell}%
+\frac{r_{+}r_{-}}{4G\ell^{2}}\frac{db}{d\ell}-\frac{r_{+}^{2}+r_{-}^{2}%
}{4G\ell^{3}}a-\frac{r_{+}r_{-}}{2G\ell^{3}}b,
\end{equation}
and
\begin{align*}
T\frac{dS}{d\ell}  &  =\frac{r_{+}^{2}-r_{-}^{2}}{4G\ell^{2}}\frac{da}{d\ell
}+\frac{r_{-}\left(  r_{+}^{2}-r_{-}^{2}\right)  }{4G\ell^{2}r_{+}}\frac
{db}{d\ell},\\
\Omega\frac{dJ}{d\ell}  &  =\frac{r_{-}^{2}}{4G\ell^{2}}\frac{da}{d\ell}%
+\frac{r_{-}\left(  r_{+}^{2}+r_{-}^{2}\right)  }{8G\ell^{2}r_{+}}\frac
{db}{d\ell}-\frac{r_{-}^{2}}{4G\ell^{3}}a-\frac{r_{-}\left(  r_{+}^{2}%
+r_{-}^{2}\right)  }{8G\ell^{3}r_{+}}b,\\
V\frac{dP}{d\ell}  &  =-\frac{r_{+}^{2}}{4G\ell^{3}}a-\frac{r_{-}\left(
3r_{+}^{2}-r_{-}^{2}\right)  }{8G\ell^{3}r_{+}}b.
\end{align*}
Thus, it is easy to see that
\begin{equation}
\frac{dM}{d\ell}\neq T\frac{dS}{d\ell}+\Omega\frac{dJ}{d\ell}+V\frac{dP}%
{d\ell},
\end{equation}
which indicates that it is impossible to find the proper thermodynamic
volume\ to maintain the first law and the homogenous Smarr formula
simultaneously if the cosmological constant is taken as the pressure
$P\equiv\frac{1}{8\pi G\ell^{2}}$. This seems that the interpretation of
enthalpy was inapplicable for the 3D AdS black hole in the general gravity models.

When we deal with the concrete gravity model, however, it is found that the
simultaneous preservation of the first law and the homogeneous Smarr formula
will give some further constraints when the parameters $a$ and $b$ are
dependent on the AdS radius. But whether the constraints are reasonable has to
be checked. Here we take the GMMG \cite{ms14} for example to study this and
start with its Lagrangian%
\begin{equation}
L_{M}=L_{G}+\frac{1}{2\mu}L_{CS}+\frac{1}{m^{2}}L_{H}+h_{a}T^{a}+\frac{\alpha
}{2}\epsilon^{abc}e_{a}h_{b}h_{c}%
\end{equation}
where $L_{CS}=\frac{1}{8\pi G}\omega_{a}\left(  d\omega^{a}+\frac{2}%
{3}\epsilon^{abc}\omega_{b}\omega_{c}\right)  $ is the topological
modification term, $L_{H}=$ $\frac{1}{8\pi G}\left(  f^{a}R_{a}+\frac{1}%
{2}\epsilon^{abc}e_{a}f_{b}f_{c}\right)  $ is the higher-order derivative
modification term, and $\mu$, $m$, $\alpha$ are the parameters which are
introduced in TMG, NMG and MMG. For GMMG model, it was obtained \cite{sa15}
\begin{equation}
a=1+\frac{\gamma}{2\mu^{2}\ell^{2}}+\frac{s}{2m^{2}\ell^{2}},\text{ \ }%
b=\frac{1}{\mu\ell},
\end{equation}
where $\gamma$, $s$ are the constants. Through a tedious calculation, it is
found that only if
\begin{equation}
\mu\ell=\text{constant and }m\ell=\text{constant}, \label{pc}%
\end{equation}
we can ensure both the first law and the homogeneous Smarr formula, which
means that the interpretation of thermodynamic enthalpy for BTZ black holes in
general gravity models enforces the extra constraints as in Eq. (\ref{pc})
when we promote the cosmological constant into a variable thermodynamic parameter.

Now we try to understand why there is such constraints as in Eq. (\ref{pc}).
First, these constraints are consistent with the previous discussions in
different 3D gravity models, in particular for some special situations, i.e.
$\mu\ell=1$ in TMG model leads to 3D chiral gravity \cite{lss08}; $m^{2}{\ell
}^{2}=\frac{1}{2}$ (or $\Lambda_{0}/m^{2}=-1$ ) in NMG leads to an extra gauge
symmetry at the linearized level which allows massive modes to become
partially massless \cite{deg13}.

On the other hand, the BTZ metric is locally isomorphic to the AdS vacuum, so
any theory of 3D gravity admitting an AdS vacuum will also admit BTZ black
holes. Now we search for the AdS vacuum for GMMG models which is also the
maximally symmetric vacuum \cite{bht092} of GMMG defined by
\begin{equation}
G_{\mu\nu}=-\Lambda g_{\mu\nu}.
\end{equation}
Thus, the GMMG field equation
\begin{equation}
G_{\mu\nu}+\Lambda_{0}g_{\mu\nu}+\frac{1}{\mu}C_{\mu\nu}+\frac{\gamma}{\mu
^{2}}J_{\mu\nu}+\frac{s}{2m^{2}}K_{\mu\nu}=0,
\end{equation}
where $G_{\mu\nu}$ is Einstein tensor, $C_{\mu\nu}=$ $\frac{1}{\sqrt{-g}%
}\varepsilon_{\mu\alpha\beta}\triangledown^{\alpha}\left(  R_{\nu}^{\beta
}-\frac{1}{4}\delta_{\nu}^{\beta}R\right)  $ is Cotton tensor, and $J_{\mu\nu
}$, $K_{\mu\nu}$ are the higher-order modification terms (see Ref.
\cite{lss08} for their expressions), will be reduced to%
\begin{equation}
\Lambda_{0}-\Lambda+\frac{\gamma\Lambda^{2}}{4\mu^{2}}-\frac{s\Lambda^{2}%
}{4m^{2}}=0,
\end{equation}
which solves the effective cosmological constant%
\begin{equation}
\Lambda=\frac{2\left(  1\pm\sqrt{1-\Lambda_{0}\left(  \frac{\gamma}{\mu^{2}%
}-\frac{s}{m^{2}}\right)  }\right)  }{\frac{\gamma}{\mu^{2}}-\frac{s}{m^{2}}}.
\label{cc}%
\end{equation}
It is consistent with that obtained in the linearization of GMMG field
equation around the AdS metric \cite{sa15}. The AdS space requires the
effective cosmological constant $\Lambda=-\frac{1}{\ell^{2}}<0$. This enforces
the coupling parameters to satisfy certain conditions, i.e. the minus sign for
the sign \textquotedblleft$\pm$\textquotedblright\ can be taken but must
$\frac{\gamma}{\mu^{2}}-\frac{s}{m^{2}}>0,\Lambda_{0}<0$. Thus, we find the
AdS vacuum for GMMG models with $\Lambda=-\frac{1}{\ell^{2}}$, which means
that we must also find BTZ black holes because these are locally isometric to
the AdS vacuum.

On the other hand, Ref. \cite{krt09} has extended the Komar integral relation
for the asymptotically flat black holes to that for the asymptotically AdS
black holes,
\begin{equation}
\frac{1}{8\pi G}\int_{\partial\Sigma}dS_{ab}\left(  \nabla^{a}\xi^{b}%
+2\Lambda\omega^{ab}\right)  =0, \label{ki}%
\end{equation}
where $dS_{ab}$ is the volume element normal to the co-dimension 2 surface
$\partial\Sigma$ which is the boundary of the hypersurface $\Sigma$ in BTZ
black hole spacetime, $\xi^{a}$ is the Killing vector on this spacetime, and
$\omega^{ab}$ is the anti-symmetric Killing potential by solving $\xi
^{b}=\nabla_{a}\omega^{ab}$. For the boundary $\partial\Sigma=\partial
\Sigma_{\infty}\cap\partial\Sigma_{r_{+}}$, one can rearrange the Komar
integral,%
\begin{equation}
\frac{1}{8\pi G}\int_{\partial\Sigma_{\infty}}dS_{ab}\left(  \nabla^{a}\xi
^{b}+2\Lambda\omega^{ab}\right)  =\frac{1}{8\pi G}\int_{\partial\Sigma_{r_{+}%
}}dS_{ab}\left(  \nabla^{a}\xi^{b}+2\Lambda\omega^{ab}\right)  ,
\end{equation}
which leads to the homogeneous Smarr formula directly for normal BTZ black
holes \cite{fmm15}. However, in general 3D gravity models, when we promote the
cosmological constant $\Lambda$ into a variable, those coupling constants,
i.e. $\mu$, $m$, will also become variables, as seen from Eq. (\ref{cc}). In
order to ensure both the thermodynamic first law and the homogeneous Smarr
formula to hold, one must constraint these coupling parameters to satisfy the
relation in Eq. (\ref{pc}), which might be the justification for these constraints.

Moreover, from the left and right central charges of dual CFT \cite{ms14} of
GMMG model,
\begin{align}
C_{L}  &  =\frac{3\ell}{2G}\left(  1-\frac{1}{\mu\ell}+\frac{\gamma}{2\mu
^{2}\ell^{2}}+\frac{s}{2m^{2}\ell^{2}}\right)  ,\nonumber\\
C_{R}  &  =\frac{3\ell}{2G}\left(  1+\frac{1}{\mu\ell}+\frac{\gamma}{2\mu
^{2}\ell^{2}}+\frac{s}{2m^{2}\ell^{2}}\right)  ,
\end{align}
the microscopic Cardy formula leads to the entropy \cite{lss08,as98,ss00}%
\begin{equation}
S=\frac{\pi^{2}\ell}{3}\left(  C_{L}T_{L}+C_{R}T_{R}\right)  \label{cfe}%
\end{equation}
where $T_{L}=\frac{r_{+}-r_{-}}{2\pi\ell^{2}},T_{R}=\frac{r_{+}+r_{-}}%
{2\pi\ell^{2}}$ are the left and right temperatures of BTZ black holes
respectively. Thus, we see that when the condition in Eq. (\ref{pc}) holds,
the entropy (\ref{cfe}) can still be regarded as independent on the AdS
radius, even if the cosmological constant is promoted to a variable. This is
also consistent with the requirement of differential expression for
thermodynamic first law, in which the pressure $P$ and the entropy $S$ should
be considered as independent variables to each other.

Finally, there exists the continuous phase transition for BTZ black holes in
general 3D gravity models. This is different from that in the normal Einstein
gravity, in which BTZ black holes exhibits no interesting phase behaviour
\cite{fmm15}, since its heat capacity is always positive as verified by the
method of thermodynamic curvature \cite{cc99}. We still take GMMG models for
example. Considering the constraints (\ref{pc}), the heat capacity is
calculated as%
\[
C=\frac{\partial M}{\partial T}|_{P,J}=\frac{4\pi\left(  a^{2}-b^{2}\right)
r_{+}\left(  r_{+}^{2}-r_{-}^{2}\right)  \ell}{br_{-}\left(  3r_{+}^{2}%
+r_{-}^{2}\right)  \ell+ar_{+}\left(  r_{+}^{2}+3r_{-}^{2}\right)  },
\]
where $8G=1$ is taken for the presentation in Fig.1. It is noted that for our case, only if the
parameters $a$ and $b$ have different signs, the thermodynamic process gives
the interesting phase transition.

\begin{figure}[ptb]
\centering
\includegraphics[width=4.75in]{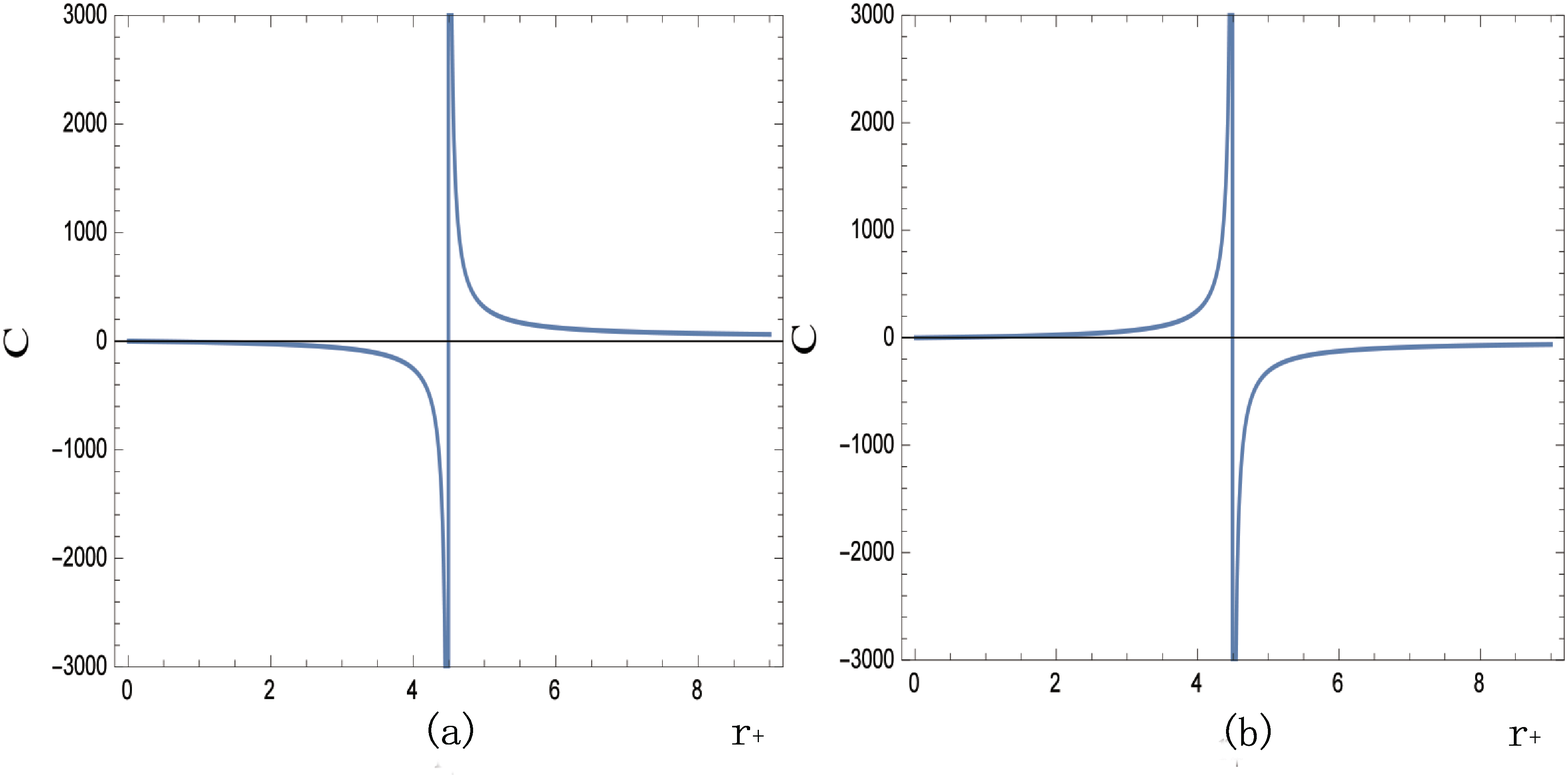} \caption{The heat capacity $C$ as
functions $r_{+}$ for BTZ black holes in GMMG model, with $\ell=1, \text{
\ }r_{-}=0.1$, while $a=-2, \text{ \ }b=3$ for (a) and $a=2, \text{ \ }b=-3$
for (b). }%
\label{fig1}%
\end{figure}
\ \ \ \ \ \ \ \ \ \ \ \ \ \ \ \ \ \ \ \ \ \ \ \ \ \ \ \ \ \ \ \ \ \ \ \ \ \ \ \ \ \ \ \ \ \ \ \ \ \ \ \ \ \ \ \ \ \ \ \ \ \ \ \ \ \ \ \ \ \ \ \ \ \ \ \ \ \ \ \ \ \ \

\section{Conclusion}

In this paper, we have investigated homogeneous Smarr relation and the
interpretation of thermodynamic enthalpy for the mass of BTZ black holes under
the background of 3D gravity theory. For normal 3D BTZ black holes, it is
direct to reduce the usual relations of Eqs. (\ref{1fl}) and (\ref{hsf}) to 3D
to obtain the corresponding Smarr formula. For exotic BTZ black holes, it is
found that the thermodynamic volume is not geometric and dependent on the
angular momentum and the pressure, different from the situation of normal BTZ
black holes. However, when we extend these into BTZ black holes in the
background of general 3D gravity models, it is impossible to find the proper
conjugate thermodynamic volume when considering the variable cosmological
constant as the pressure, unless some extra conditions are added. We have
studied these conditions for GMMG model, and found that the conditions in Eq.
(\ref{pc}) not only ensure both the first law and Smarr formula to hold
simultaneously, but also accord with all previous discussions for GMMG model
where AdS radius is not regarded as a variable. In particular, the constraint
conditions are consistent with the phenomena of continuous phase transition
occurred for BTZ black holes in general 3D gravity models.

\section{Acknowledge}

The authors would like to thank the support by Grant Nos. 11374330 and
91636213 of the National Natural Science Foundation of China and by the
Fundamental Research Funds for the Central Universities, China University of
Geosciences (Wuhan) (No. CUG150630).

\end{document}